\def\year{2020}\relax
\documentclass[letterpaper]{article} 
\usepackage{aaai20}  
\usepackage{times}  
\usepackage{helvet} 
\usepackage{courier}  
\usepackage[hyphens]{url}  
\usepackage{graphicx} 
\usepackage{graphicx}  
\usepackage{amssymb}
\usepackage{amsmath}
\usepackage{multirow}
\usepackage{subcaption}

\title{Towards Comprehensive Recommender Systems: Time-Aware Unified Recommendations Based on Listwise Ranking of Implicit Cross-Network Data}
\author{\Large \textbf{Dilruk Perera} and Roger Zimmermann\\
\Large School of Computing, National University of Singapore \\ 
\Large dilruk@comp.nus.edu.sg and rogerz@comp.nus.edu.sg
\\ 
}

\author{\Large \textbf{Dilruk Perera} and Roger Zimmermann\\
\Large School of Computing, National University of Singapore \\ 
\Large dilruk@comp.nus.edu.sg and rogerz@comp.nus.edu.sg
\\ 
}
 
\begin{document}

\maketitle

\begin{abstract}
The abundance of information in web applications make recommendation essential for users as well as applications. Despite the effectiveness of existing recommender systems, we find two major limitations that reduce their overall performance: (1) inability to provide timely recommendations for both new and existing users by considering the dynamic nature of user preferences, and (2) not fully optimized for the ranking task when using implicit feedback. Therefore, we propose a novel deep learning based unified cross-network solution to mitigate cold-start and data sparsity issues and provide timely recommendations for new and existing users. Furthermore, we consider the ranking problem under implicit feedback as a classification task, and propose a generic personalized listwise optimization criterion for implicit data to effectively rank a list of items. We illustrate our cross-network model using Twitter auxiliary information for recommendations on YouTube target network. Extensive comparisons against multiple time aware and cross-network baselines show that the proposed solution is superior in terms of accuracy, novelty and diversity. Furthermore, experiments conducted on the popular MovieLens dataset suggest that the proposed listwise ranking method outperforms existing state-of-the-art ranking techniques. 
\end{abstract}

\section{Introduction}
Recommender systems have been effectively used in various web applications (e.g., search engines, e-commerce and social networks) to mitigate the notorious information overload problem. In addition to the user benefits, where interesting information are automatically recommended to users, recommender systems help web applications to significantly increase their revenue. For example, 35\% of Amazon purchases and 75\% of Netflix movie views are initiated by their internal recommender systems. Therefore, over the years, constant efforts have been made to improve the quality of recommendations in terms of accuracy, novelty and diversity \cite{perera2018lstm,mclaughlin2004collaborative}. However, despite their success, existing systems face two main limitations that continue to hinder their performances. 

(1) \textbf{Inability to provide timely recommendations for new and existing users:} Due to the high item-to-user ratio in web applications, existing user interactions are highly sparse, and leads to low recommender performance (\textit{data sparsity problem}). It is also infeasible to determine new user preferences due to the absence of their interactions (\textit{cold-start problem}). Moreover, user preferences toward items are also subject to constant change over time \cite{campos2012performance,koren2009collaborative}. Therefore, recommender systems should capture such changes and model up-to-date user preferences for effective recommendations for both new and existing users.

(2) \textbf{Not fully optimized to use implicit feedback data for listwise ranking:} Unlike interactions with explicit feedback (1-5 ratings), most user interactions are implicit in nature (e.g., watching a video or clicking on a link). Therefore, recent years have seen a shift in recommender systems from using explicit ratings to widely and cheaply available implicit ratings \cite{kelly2003implicit}. In practice, recommender systems output a ranked list of items, based on user preferences. However, existing implicit feedback based systems, for example, Matrix Factorization (MF) \cite{koren2009matrix}, Factorization Machines (FM) \cite{rendle2010factorization} and vast majority of deep learning models \cite{he2017neuralfact} are not fully optimized to rank a list of items. They are commonly optimized for predicting user preferences for a single item (\textit{pointwise}) or, ranking preferences for a pair of items (\textit{pairwise}). Therefore, the achievable recommender quality is limited.

We propose a time aware unified cross-network solution with a novel personalized listwise loss function to address the above limitations and create a consolidated recommender solution. User engagements on multiple social networks showcase their preferences from various perspectives. Therefore, such preferences across networks can be used to mitigate cold start and data sparsity issues, while improving recommender novelty and diversity. Therefore, for new users, at the time of registration, their preferences are obtained from other source networks. For existing users, more comprehensive preferences are obtained by integrating their preferences with other source networks. 

To capture the dynamic nature of user preferences, we modeled preferences under three temporal levels, namely short, long and long short term preferences. Short term preferences reflect most recent preferences and are highly likely to drive immediate future interactions. Long term preferences are practised over an extended period and reflect general preferences. Unlike short term preferences, long term preferences are less affected by the current context and recent trends. However short term preferences are highly sensitive to preference outliers. For example, a user may be drawn to watch athletics videos during the hype of the Olympics season. Once the Olympics is over, he may no longer be interested in watching such videos. Therefore, we introduced a novel long short term preference component to effectively remove such outliers. The main intuition is to exploit historical user preferences that are similar to the current preferences, and compute a new (long short term) preference component based on the preference similarities and frequencies, which is less affected by the current outliers and highly relevant for immediate user interactions.

\noindent We summarize our main contributions as follows:
\begin{itemize}
	\item We proposed a novel deep learning based time aware unified cross-network model that captures user preferences in three temporal levels for effective recommendations.
	\item We consider the ranking problem under implicit feedback as a classification task, and proposed a novel, generic listwise optimization criterion.
	\item We conducted extensive experiments to demonstrate that the proposed model and optimization criterion consistently outperform the state-of-the-art baselines.
\end{itemize}

\section{Literature Review}
User preferences from multiple networks have increased the robustness of recommendations against cold start and data sparsity issues \cite{mehta2005ontologically,yan2015unified}. For example, YouTube recommender accuracy was increased using information from Google+ \cite{sang2015cross}, Twitter \cite{wanave2016youtube}, and Amazon recommender accuracy was increased with information from Cheetah Mobile \cite{hu2018conet}. However, most cross-network solutions are non-unified and solve either cold-start or data sparsity issues. 

Recently, a handful of efforts have been made to utilize auxiliary information from source networks to develop unified recommender solutions \cite{yan2015unified,yan2016unified}. However, these solutions do not model the dynamic nature of user preferences and recommender performance degrades over time unless the models are retrained frequently. To the best of our knowledge, the first time aware unified solution was proposed in \cite{perera2017exploring}, where the authors used a decaying function to prioritize most recent preferences when conducting recommendations. However, the model is based on a linear MF model and optimized using a pointwise criterion.

Widely used pointwise optimization criteria are less effective for implicit feedback based approaches due to the absence of negative feedback. As a solution, Bayesian Personalized Ranking (BPR) was proposed to maximize the difference between predicted ratings for interacted and non-interacted items \cite{rendle2009bpr}. However, BPR and its variants \cite{yu2018multiple,loni2016bayesian}, were not fully optimized to rank a list of items. Similar work on learning to rank with non-collaborative models have also been proposed (e.g., modeling distributions on permutations \cite{kondor2007multi,huang2008efficient}. Nevertheless, these solutions are not personalized and are not well suited for the recommendation task. Listwise optimization approaches based on explicit feedback showed superiority over pointwise and pairwise approaches for recommending a ranked list of items \cite{cao2007learning}.

\section{Model Preliminaries}
\subsubsection{Intuition for the Listwise Loss:} 
\begin{figure*}
    \centering
    \includegraphics[width=\linewidth]{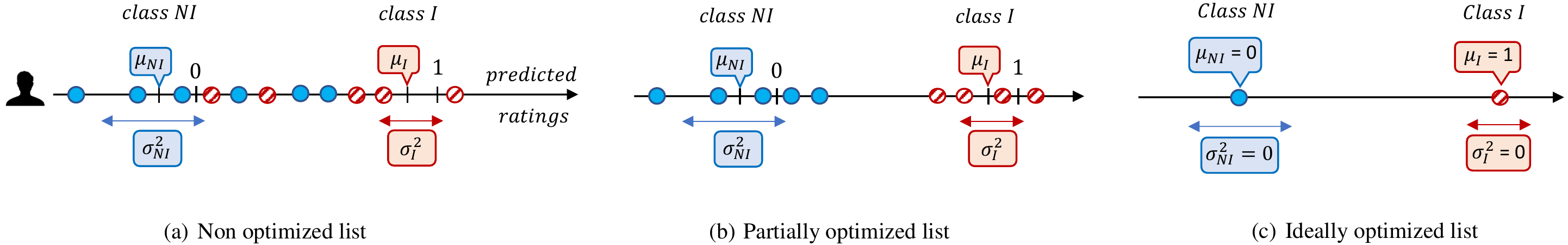}
    \caption{Stages of the listwise loss optimization process.}
    \label{fig:opt_criterion}
\end{figure*}

Given a list of items with corresponding implicit feedback given by a user, the optimization criterion aims to classify item feedback into 2 classes (interacted and non-interacted). The goal is to maintain high inter-class distances and low intra-class distances. Considering ground truth values for interacted and non-interacted feedback as 1 and 0, the optimization criterion should lead the mean values of the corresponding  classes ($\mu_{I}$ and $\mu_{NI}$) to 1 and 0, and the variance values of both classes ($\sigma_{I}^2$ and $\sigma_{NI}^2$) to 0. Accordingly, a non optimized list of items will not have clearly separated classes, a partially optimized list will have a clear separation, and an ideally optimized list will have all values of each class merged into a single point (see Figure \ref{fig:opt_criterion}). Note that, the proposed personalized loss function is generic and can be used with popular collaborative filtering approaches (e.g. MF, FM and kNN). 

\subsubsection{Pairwise Loss verses Listwise Loss:} Unlike explicit feedback, implicit feedback lacks the notion of preference levels due to its binary nature. Therefore, popular pointwise loss functions that are optimized to predict preference levels are ineffective for implicit feedback. Pairwise loss is the widely adopted optimization criterion for learning with implicit feedback. Although recommendation is widely considered to be a ranking task, we argue that under implicit feedback, it is a binary classification task. Considering the list of items for a user, the aim is to classify the items into 2 classes (interacted and non-interacted). The pairwise approach is not suitable for this task, since it is only optimized to rank a pair of items, and the ranking for the item list is indirectly obtained through transitivity ($i_1 > i_2$, and $i_2 > i_3$ means $i_1 > i_3$). In the proposed listwise loss, the optimization is more realistic in nature as it directly optimizes for the entire list in each training instance.

It is also infeasible to run pairwise optimizations for all item pairs, since it requires $^nC_2$ ($n$ is the no. of items) runs per user, per training epoch. Hence, pairwise optimization is executed on a randomly selected subset of item pairs for each user, at each instance by assuming that multiple executions would generalize to rank the entire item list. Thus, effectiveness largely depends on the quality of the randomly selected item pairs, and no of pairs. Also, since ranking is obtained through transitivity, item predictions need to be computed multiple times to obtain item list rankings. For example, to obtain $i_1 > i_3$, from $i_1 > i_2$ and $i_2 > i_3$, prediction for $i_2$ need to be computed in two runs. This is highly inefficient considering the number of items in applications. In contrast, the listwise approach only computes the prediction for an item once and has a similar complexity to the inferior pointwise approach. Furthermore, since pairwise approach does not compute ratings for the entire list in a single run, during validation and testing, predictions for each user-item pair need to be computed in different runs and ranked in a separate process. In the listwise approach, the model is already trained to provide a classified list for testing.

\subsubsection{Problem Formulation:}
We denoted each existing user on the target network at a time interval $t$ as $ u_e^t \in U = [{Sr}^t_{e},{Tg}^t_{e}]$, using his source and target network interaction histories (${Sr}^t_{e}$ and ${Tg}^t_{e}$) over $T = \{ 1, ..., t \}$ time intervals. Similarly, each new user is denoted as $u_n^t \in U = [{Sr}^t_{n}]$, using only source network interactions (${Sr}^t_{n}$), since new user interactions are unavailable on the target network. Given the interaction histories, at the end of each interval $t$, the goal is to determine their preferences and predict future interactions at $t+1$, on the target network.

\section{Methodology}
The proposed model conducts recommendations for new and existing users over four stages (see Figure \ref{fig:main_archi}). First, historical interactions are transferred and integrated in the cross-network topical layer. Second, three levels of user preferences are extracted from the computed topical representations on both source and target networks.  Third, extracted preferences are integrated to obtain overall preferences of new and existing users. Fourth, the overall preferences are used to conduct unified recommendations.

\subsection{Cross-Network Topical Layer}
User interactions on different networks are often multi-modal and heterogeneous in nature (e.g., tweets and liked YouTube videos). Since comparison of heterogeneous interactions across networks is challenging, we used topic modeling to transform user interactions onto a homogeneous topical space (see Figure \ref{fig:main_archi}, stage 1). We assumed that each user interaction is associated with multiple topics and extracted related topics from textual metadata (i.e., tweet contents, video titles and descriptions). We considered each tweet and interacted YouTube video as a \textit{document} and extracted topics using Twitter-Latent Dirichlet Allocation (Twitter-LDA) \cite{zhao2011comparing}, which is highly effective against short and noisy textual contents.  

Consequently, each interaction is represented as a topical distribution over all possible topics. Source network interactions of new users within a given time interval $t$ are represented as $\boldsymbol{x_{Sr}^{t}} = \{f_{Sr}^{t,1}, \dots, f_{Sr}^{t, K^t}\} \in \mathbb{R}^{K^t}$, the summation over all topical distributions during $t$, where $K^t$ is the number of topics. Each element $f_{Sr}^{t, k^t} \in \boldsymbol{x_{Sr}^{t}}$ in the resulting topical distribution is a topical frequency representing user preference level towards the corresponding topic. Therefore, topical distributions are a good representation of user preferences. Similarly, existing users within time interval $t$ are represented as $\{\boldsymbol{x_{Sr}^{t}};\boldsymbol{x_{Tg}^{t}}\} \in \mathbb{R}^{2K^t}$, using source and target network topical distributions. Thus, new and existing users are represented using their interaction histories over $T$ intervals as $u_n = \{\boldsymbol{x_{Sr}^{1}},\dots, \boldsymbol{x_{Sr}^{t}}\} \in \mathbb{R}^{T\times K^t}$ and $u_e = \{\boldsymbol{x_{Tg}^{1}},\dots, \boldsymbol{x_{Tg}^{t}}; \boldsymbol{x_{Sr}^{1}},\dots, \boldsymbol{x_{Sr}^{t}}\} \in \mathbb{R}^{T\times 2K^t}$, which form the inputs to the model. 

\subsection{User Preference Extraction}
\begin{figure*}
    \centering
    \includegraphics[width=0.9\linewidth]{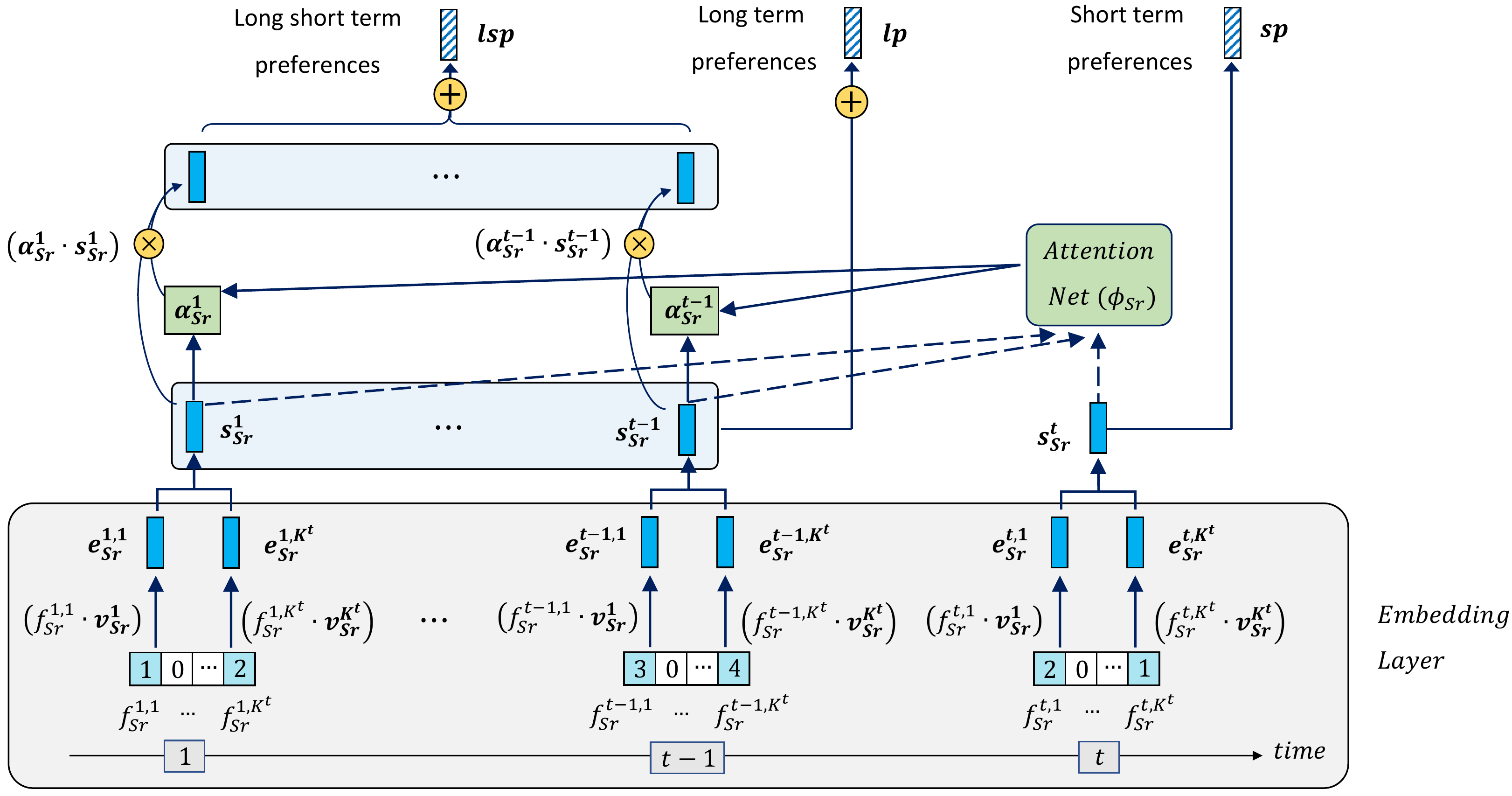}
    \caption{Overview of the user preference extraction process on the source network, which computes short, long and long short term user preferences. Note that, this is analogous to the extraction process on the target network.}
    \label{fig:user_pref_computation}
\end{figure*}

We extracted user preferences as short, long and long short term preferences (see Figure \ref{fig:main_archi}, stage 2). This extraction process is further illustrated in Figure \ref{fig:user_pref_computation}.
\subsubsection{Embedding Layer:} Given a source network topical distribution $\boldsymbol{x_{Sr}^{t}} = \{f_{Sr}^{t,1},\dots, f_{Sr}^{t, K^t}\} \in \mathbb{R}^{K^t}$, the embedding layer projects each input topic to a dense vector representation and computes its corresponding embeddings as $E_{Sr}^{t} = \{\boldsymbol{e_{Sr}^{t,1}},\dots, \boldsymbol{e_{Sr}^{t, K^t}}\} \in \mathbb{R}^{K^t \times k}$, where $\boldsymbol{e_{Sr}^{t,c}} = f_{Sr}^{t,c} \cdot \boldsymbol{v_{Sr}^c} \in \mathbb{R}^k$ is the embedding for topic $c$, $f_{Sr}^{t,c}$ is the topical frequency, $\boldsymbol{v_{Sr}^c}$ is the learnt global vector representation of topic $c$, and $k$ is the dimensionality of the embedding space. We only learnt embeddings for non-zero topical frequencies since $f_{Sr}^{t,c}=0$ leads to $\boldsymbol{e_{Sr}^{t,c}}=0$. Similarly, a separate set of vector representations $\boldsymbol{v_{Tg}^c}$ are learnt for topics on the target network. Therefore, for $n$ and $m$ non-zero frequencies on source and target networks, resulting embedding vectors are $E_{Sr}^{t} \in \mathbb{R}^{k \times n}$ and $E_{Tg}^{t} \in \mathbb{R}^{k \times m}$. The embeddings are used to compute short, long and long short term user preferences.
The intuition for learning separate sets of topical vector representations for each network is to distinctly capture the network-level topical properties. For example, entertainment related topics are more prominent on YouTube than Twitter \cite{perera2017exploring}. Therefore, the same topic will be differently represented on each network.

\subsubsection{Short Term Preferences:} We extracted short term preferences from interactions within the most recently completed time interval as it is the closest representation of the current context. At the end of time interval $t$, short term preferences of new users are computed using source network embeddings within the time interval \big($E_{Sr}^{t} \in \mathbb{R}^{k\times n}$\big), and short term preferences of existing users are computed using both source and target network embeddings \big($E_{Sr}^{t} \in \mathbb{R}^{k\times n}$ and $E_{Tg}^{t} \in \mathbb{R}^{k\times m}$\big) as follows: 
\begin{align}
    \boldsymbol{sp_{n}^{t}} & = \{\boldsymbol{s_{Sr}^{t}}\} = \sum_{c\in C_n} \boldsymbol{e_{Sr}^{t,c}} & \in \mathbb{R}^{k}\\
    \boldsymbol{sp_{e}^{t}} & = \{\boldsymbol{s_{Sr}^{t}}; \boldsymbol{s_{Tg}^{t}}\},\nonumber \\ 
                           & = \sum_{c\in C_n} \boldsymbol{e_{Sr}^{t,c}} \ ; \ \sum_{c\in C_m} \boldsymbol{e_{Tg}^{t,c}} & \in \mathbb{R}^{2k}
\end{align}
\noindent where $\boldsymbol{sp_{n}^{t}}$ and $\boldsymbol{sp_{e}^{t}}$ are short term preferences of new and existing users, $C_n$ and $C_m$ are sets of indices of the non-zero topical frequencies in source and target networks.

\subsubsection{Long Term Preferences:} Long term preferences are modeled using the collection of short term preferences over time. Hence, at the end of time interval $t$, long term preferences for new and existing users are computed as follows:
\begin{align}
        \boldsymbol{lp_{n}^{t}} & = \sum_{t=1}^{t-1} \boldsymbol{s_{Sr}^{t}} & \in \mathbb{R}^k \\
        \boldsymbol{lp_{e}^{t}} & = \sum_{t=1}^{t-1} \boldsymbol{s_{Sr}^{t}} \ ; \ \sum_{t=1}^{t-1}\boldsymbol{s_{Tg}^{t}} & \in \mathbb{R}^{2k}
\end{align}
\noindent where $\boldsymbol{lp_{n}^{t}}$ and $\boldsymbol{lp_{e}^{t}}$ are long term preferences of new and existing users.

\subsubsection{Long Short Term Preferences:} We used a neural network based attention mechanism to compute long short term preferences as a weighted sum of long term preferences. Specifically, the neural network computes attention scores (i.e., similarity based on relevancy and frequency) between past and current preferences. Thereby, only the relevant past preferences are used to compute the long short term preferences which reduces the outlier effects on short term preferences. Thus, for each user, given a pair of current and past preference embeddings on source ($\boldsymbol{s_{Sr}^{t}}$ and $\boldsymbol{s_{Sr}^{t'}}$) or target networks ($\boldsymbol{s_{Tg}^{t}}$ and $\boldsymbol{s_{Tg}^{t'}}$), the attention scores are computed as follows:
\begin{align}
    \alpha^{t'}_{Sr} = \phi_{Sr} (\boldsymbol{s_{Sr}^{t}}, \boldsymbol{s_{Sr}^{t'}})\\
    \alpha^{t'}_{Tg} = \phi_{Tg} (\boldsymbol{s_{Tg}^{t}}, \boldsymbol{s_{Tg}^{t'}})
\end{align}
\noindent where $\phi_{Sr}$ and $\phi_{Tg}$ are neural attention functions for source and target networks, $\alpha^{t'}_{Sr} \in \mathbb{R}$ and $\alpha^{t'}_{Tg} \in \mathbb{R}$ are attention scores between past ($t'$) and current ($t$) preferences. Once attention scores are computed for all past intervals, long term preferences are weighted as follows:
\begin{align}
    \boldsymbol{lsp_{n}^t} & = \sum_{t'=1}^{t-1} \alpha_{Sr}^{t'} \cdot \boldsymbol{s_{Sr}^{t'}} \in \mathbb{R}^k\\
    \boldsymbol{lsp_{e}^t} & = \sum_{t'=1}^{t-1} \alpha_{Sr}^{t'} \cdot \boldsymbol{s_{Sr}^{t'}} \;\ \sum_{t'=1}^{t-1} \alpha_{Tg}^{{t'}} \cdot \boldsymbol{s_{Tg}^{t'}} \in \mathbb{R}^{2k}
\end{align}
\noindent where $\boldsymbol{lsp_{n}^t}$ and $\boldsymbol{lsp_{e}^t}$ are long short term preferences of new and existing users.
\begin{figure}
    \centering
    \includegraphics[width=\linewidth]{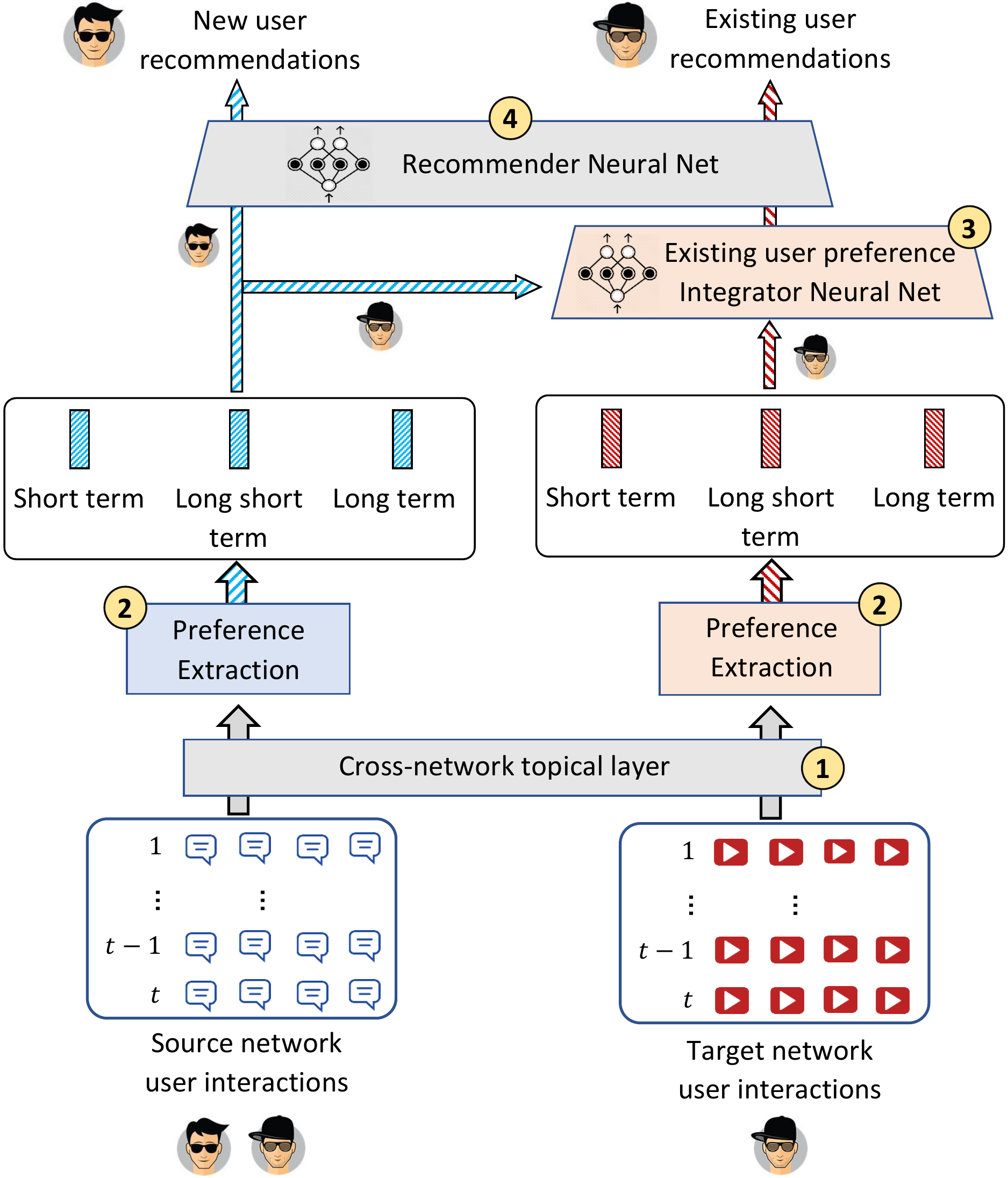}
    \caption{High-level overview of the recommendation process for new and existing users.}
    \label{fig:main_archi}
\end{figure}
\subsection{User Preference Integration}
We integrated short, long and long short term preferences of new and existing users to obtain a final preference representation that reflects their near future preferences.

\subsubsection{New User Preferences:} The temporal preferences for new users are captured from the same network space. Therefore, the final new user preference representation at the end of time interval $t$ ($\boldsymbol{{p}_n^t}$) is obtained using a direct summation over the computed preference components as follows:
\begin{equation}
    \boldsymbol{{p}_n^t} = \boldsymbol{sp_{n}^{t}} + \boldsymbol{lp_{n}^{t}} + \boldsymbol{lsp_{n}^{t}} \ \quad \in \mathbb{R}^{k}
\end{equation}

\subsubsection{Existing User Preferences:} Unlike for new users, the temporal preferences for existing users are captured on heterogeneous network spaces. Therefore, we introduced a neural integration function to effectively integrate these user preferences across networks (see Figure \ref{fig:main_archi}, stage 3). Hence, the final preference representation for existing users at the end of time interval $t$ ($\boldsymbol{p_e^t}$) is computed as follows:
\begin{equation}
    \boldsymbol{p_e^t} = \Phi_{E} (\boldsymbol{v_{e}}, \boldsymbol{sp_{e}^{t}}, \boldsymbol{lp_{e}^{t}}, \boldsymbol{lsp_{e}^{t}}) \ \quad \in \mathbb{R}^{k}
\end{equation}
\noindent where $\Phi_{E}$ is the neural integration function. We also introduced an additional latent embedding $\boldsymbol{v_{e}} \in \mathbb{R}^{D'}$ to represent unique characteristics of each existing user (e.g., rating styles) to further support personalized recommendations. For example, some users may normally give higher ratings when they enjoy items, others may have high expectations and tend to give lower ratings on average. Hence, these highly personalized characteristics can be latently encoded in their embeddings.

\subsection{Recommendation} 
Many early recommender solutions modeled linear relationships between user-item interactions (e.g., MF and FM). However, recent neural network based recommender solutions showcased that neural functions better model complex user-item interactions \cite{he2017neural}. Therefore, we used a feed forward neural network (see Figure \ref{fig:main_archi}, stage 4) to model user-item pairs and obtain rating predictions as follows: 
\begin{align}
    \hat{r}_{ni}^{t+1} & = \Phi_{R} (\boldsymbol{p_{n}^t}, \boldsymbol{v_{i}}) \in \mathbb{R}\\
    \hat{r}_{ei}^{t+1} & = \Phi_{R} (\boldsymbol{p_{e}^t}, \boldsymbol{v_{i}}) \in \mathbb{R}
\end{align}
\noindent where $\hat{r}_{n,i}^{t+1}$ and $\hat{r}_{e,i}^{t+1}$ are predicted ratings for new and existing users for an item $i$ during $t+1$, $\boldsymbol{v_i} \in \mathbb{R}^{D}$ is the latent item embedding which represents the unique characteristics of each item $i$, and $\Phi_R$ is the neural recommender function.

\subsection{Optimization}
The optimization function contains a novel listwise loss and an attention loss component.

\subsubsection{Listwise Loss Component:} The model is trained at the end of each time interval $t$, using listwise training instances. These instances contain actual (ground truth) interacted and non-interacted items at the next interval $t+1$. Hence, we define each listwise training instance for a new user $u_{n}^{t}$ at $t$ as a triplet $H_n^{t} = \{(u_{n}^{t}, I_{n+}^{t+1}, I_{n-}^{t+1})|u_{n}^{t}\in U_{N+}^{t+1}\land I_{n-}^{t+1}\in I\setminus I_{n+}^{t+1} \}$, where $U_{N+}^{t+1}$ is the set of new users with at least one interaction at $t+1$, $I$ denotes the set of all items, and $I_{n+}^{t+1}$ is the set of interacted items. Analogously, each listwise training instance for an existing user $u_{e}^{t}$ is denoted as $H_e^{t} = \{(u_{e}^{t}, I_{e+}^{t+1}, I_{e-}^{t+1})|u_{e}^{t}\in U_{E+}^{t+1}\land I_{e-}^{t+1}\in I\setminus I_{e+}^{t+1} \}$.

According to the classification approach, during the training process, the recommender should assign rating values of 1 for interacted items ($I_{n+}^{t+1}$ and $I_{e+}^{t+1}$) and 0 for non-interacted items ($I_{n-}^{t+1}$ and $I_{e-}^{t+1}$). Hence, the listwise loss functions to be minimized for new ($L_{lw,n}$) and existing ($L_{lw,e}$) user predictions are defined as follows:
\begin{align}
    L_{lw,n} & = \sum_{u_n^t}\sum_{i\in I_{n+}^{t+1}}\sum_{j \in I_{n-}^{t+1}}\big(1-\mu_{ni}\big)^2 + \mu_{nj}^2 + \sigma^2_{ni} + \sigma^2_{nj} \nonumber\\
    L_{lw,e} & = \sum_{u_e^t}\sum_{i\in I_{e+}^{t+1}}\sum_{j \in I_{e-}^{t+1}}\big(1-\mu_{ei}\big)^2 + \mu_{ej}^2 + \sigma^2_{ei} + \sigma^2_{ej}\nonumber
\end{align}
\noindent where $\mu_{ni}$ and $\mu_{nj}$ ($\mu_{ei}$ and $\mu_{ej}$) are mean predicted ratings of interacted and non-interacted items for new (existing) users, and $\sigma^2_{ni}$ and $\sigma^2_{nj}$ ($\sigma^2_{ei}$ and $\sigma^2_{ej}$) are corresponding variance values for new (existing) users.

\subsubsection{Attention Loss Component:} The attention neural functions for both source and target networks compute attention scores (similarity scores of [0,1]) between pairs of past and current preferences. Therefore, we introduced attention loss components to better train the attention neural networks ($\phi_{Sr}$ and $\phi_{Tg}$). During each training instance, the computed current preference of a user on a given network is duplicated to create an extra pair of inputs to the attention function of the corresponding network. Thus, a well trained attention function should output 1 to indicate the perfect match between the duplicated preferences. Thereby, the attention loss components to be minimized are defined as follows:
\begin{align}
    L_{at,n} = &\big(1- \phi_{Sr} (\boldsymbol{s_{Sr}^{t}}, \boldsymbol{s_{Sr}^{t}}) \big)^2 \\
    L_{at,e} = &\nonumber \big(1- \phi_{Sr} (\boldsymbol{s_{Sr}^{t}}, \boldsymbol{s_{Sr}^{t}}) \big)^2 +\\& \big(1- \phi_{Tg} (\boldsymbol{s_{Tg}^{t}}, \boldsymbol{s_{Tg}^{t}}) \big)^2\nonumber
\end{align}
\noindent where $L_{at,n}$ and $L_{at,e}$ are attention loss components for new and existing users. Note that, the duplicated inputs are not used for recommendations since they are auxiliary inputs used to train the attention networks and are not valid inputs.

\subsubsection{Total Loss:} The effectiveness of the proposed model depends on two main tasks: (1) accurate modeling of user preferences, and (2) effective optimization of the classification task. Since the attention loss and listwise loss components represent these two tasks, the final loss functions to be optimized are formed by combining the two loss components. Accordingly, the final loss functions for new and existing users ($L_{n}$ and $L_{e}$) are as follows:
\begin{align}
    L_{n} & = L_{lw,n} + L_{at,n}\\
    L_{e} & = L_{lw,e} + L_{at,e}
\end{align}

Note that, since both tasks are equally important for the overall effectiveness, we used a simple addition to combine the two loss components. We could also exploit other methods to model the contribution from each task (e.g., learning a hyperparameter weight for the attention loss or learning a complex neural function to integrate the two components). However, the use of addition reduces the number of hyperparameters and manages the overall complexity of the model. The empirical results also proved it to be highly effective.

\section{Experiments}     
\subsection{Datasets}

We conducted experiments on two datasets, CrossNet and MovieLens\footnote{https://grouplens.org/datasets/movielens/1m/} to evaluate the proposed cross-network model and listwise optimization criterion, respectively. 

\subsubsection{CrossNet:} We extracted overlapped users on Twitter and YouTube from two public datasets \cite{yan2014mining,lim2015mytweet} and scraped timestamps of interactions from 1\textsuperscript{st} March 2015 to 29\textsuperscript{th} February 2017. We used tweets as source network interactions and YouTube videos either liked or added to playlists as target network interactions and extracted their associated textual contents (e.g., tweet contents, video titles and descriptions). In line with common practices, we filtered out users with less than 10 interactions on Twitter and YouTube. The final dataset contained 2372 users, 12,782 YouTube videos, and the sparsity of the user-video matrix was 99.62\%. 

\subsubsection{MovieLens:} We used a popular version of the movie ratings dataset with 1 million ratings from 6000 users for 4000 movies. Similar to other studies based on implicit feedback data \cite{koren2008factorization,he2017neural}, we transformed explicit ratings to implicit ratings where each user-item pair is 1 or 0 indicating an interaction or non-interaction.
\begin{table}[h]
	\small
	\centering
	\caption{Experimental settings.}
	\label{tab:expsetting}
	\begin{tabular}{|l|l|l|l|}
		\hline
		Exp. & Granularity & Training set (\%) & Testing set (\%) \\ \hline
		1 & \multirow{2}{*}{Biweekly} & \begin{tabular}[c]{@{}l@{}}First 20 (83\%)\end{tabular} & \begin{tabular}[c]{@{}l@{}}Last 2 (17\%)\end{tabular} \\ \cline{1-1} \cline{3-4} 
		2 &  & \begin{tabular}[c]{@{}l@{}}First 22 (92\%)\end{tabular} & \begin{tabular}[c]{@{}l@{}}Last 2 (8\%)\end{tabular} \\ \hline
		3 & \multirow{2}{*}{Monthly} & \begin{tabular}[c]{@{}l@{}}First 10 (83\%)\end{tabular} & \begin{tabular}[c]{@{}l@{}}Last 2 (17\%)\end{tabular} \\ \cline{1-1} \cline{3-4} 
		4 &  & \begin{tabular}[c]{@{}l@{}}First 11 (92\%)\end{tabular} & \begin{tabular}[c]{@{}l@{}}Last 1 (8\%)\end{tabular} \\ \hline
	\end{tabular}
\end{table}
\subsection{Experimental Setup}
Recommender systems are often evaluated on training and testing datasets with temporal overlaps [9]. This leads to biases in predictions as it does not reflect a realistic recommender environment. Therefore, at each time interval, we trained the proposed model only on interactions in past time intervals to predict interactions in future time intervals. We sorted all users on CrossNet based on their total number of YouTube interactions, and selected the first and second half of users as new and existing users, respectively. We ignored all YouTube interactions of new users, and designed four main experiments with two temporal granularities and train/test ratios (see Table \ref{tab:expsetting}). Note that, training for both new and existing users is conducted simultaneously. Furthermore, we randomly selected 10\% of interactions and 10\% of non-interactions from MoviLens as the test set to evaluate the proposed listwise optimization criterion.  

We formulated video recommendation as a Top-N recommender task and predicted a ranked set of $10$ videos that the user is most likely to interact with, at the next time interval. We calculated Hit Ratio (HR) \cite{he2015trirank} to evaluate the accuracy of the proposed model. Similar to other ranking approaches \cite{rendle2009bpr}, we calculated Area Under the ROC Curve (AUC) to evaluate the optimization criterion. Both HR and AUC metrics are calculated for each participating user and results are averaged across users. 

\subsection{Baselines}
We used the following time aware, single network and cross-network based baselines to evaluate the proposed model.
\begin{itemize}
	\item \textbf{TimePop}: Recommends the most popular $N$ items in the most recent time interval to all users.  
	\item \textbf{TBKNN} \cite{campos2010simple}: Time-Biased KNN computes $K$ neighbors for each user and recommends their recent interactions to the target user. Similar to the original experiments, we used multiple $K$ values from 4 to 50 and averaged the results for comparisons.
	\item \textbf{Time-LSTM} \cite{zhu2017next}: A single network-based LSTM model that computes short and long term user preferences based on timestamps of user interactions. The authors proposed three versions, and we selected the best performing model (TimeLSTM3) for comparisons.
	\item \textbf{Unified} \cite{yan2015unified}: A non-temporal cross-network model, which uses Twitter auxiliary information for new and existing user recommendations on YouTube.
	\item \textbf{TDCN} \cite{perera2017exploring}: Time Dependent Cross-Network is an MF based model, which also uses Twitter timestamped interactions as source information to conduct recommendations on YouTube.
\end{itemize}
Furthermore, we used several popular baselines to evaluate the proposed listwise loss criterion.
\begin{itemize}
	\item \textbf{Pop}: Recommends the most popular set of items to all users.
	\item \textbf{MF} \cite{koren2009matrix}: Learns low rank embeddings for users and items and conducts user-item rating predictions based on their inner product. MF is used for implicit and explicit rating predictions. We employed the widely used pointwise loss optimization for training.
	\item \textbf{BPR-MF} \cite{rendle2009bpr}: Uses a pairwise loss function for optimization. Each training instance is a triplet $(u, i, j)$, where user $u$ has interacted (not interacted) with item $i$ ($j$). The optimization maximizes the difference between predictions $(\hat{r}_{ui} - \hat{r}_{uj})$.
\end{itemize}

\subsection{Model Parameters}
Adaptive Moment Estimation (Adam) \cite{kingma2014adam} is an extension to the popular stochastic gradient descent algorithm, and is widely used for optimizations in various deep learning applications. We used Adam for training since it adaptively updates the learning rate. All neural network architectures used in the model were restricted to only one hidden layer, to manage the overall complexity and parameters. Given the size of the output layer as $L$, the size of the preceding hidden layer was set to $2 \times L$ to minimize the number of hyperparameters. All neural layers were regularized using the dropout technique, and dropout was experimentally set to 0.3 during training to prevent overfitting. We set the number of topics, $K^t$ to 64 using a standard grid search algorithm when using Twitter-LDA to project user interactions to a cross-network topical space. The model performance was not highly sensitive to small changes in $K^t$($\approx\pm5$).

\section{Discussion}
\subsection{Model Evaluation}
\subsubsection{Prediction Accuracy:} 

\begin{figure}
    \captionsetup[subfigure]{justification=centering}
    \centering
    \begin{subfigure}[b]{0.7\linewidth}
        \includegraphics[width=\textwidth]{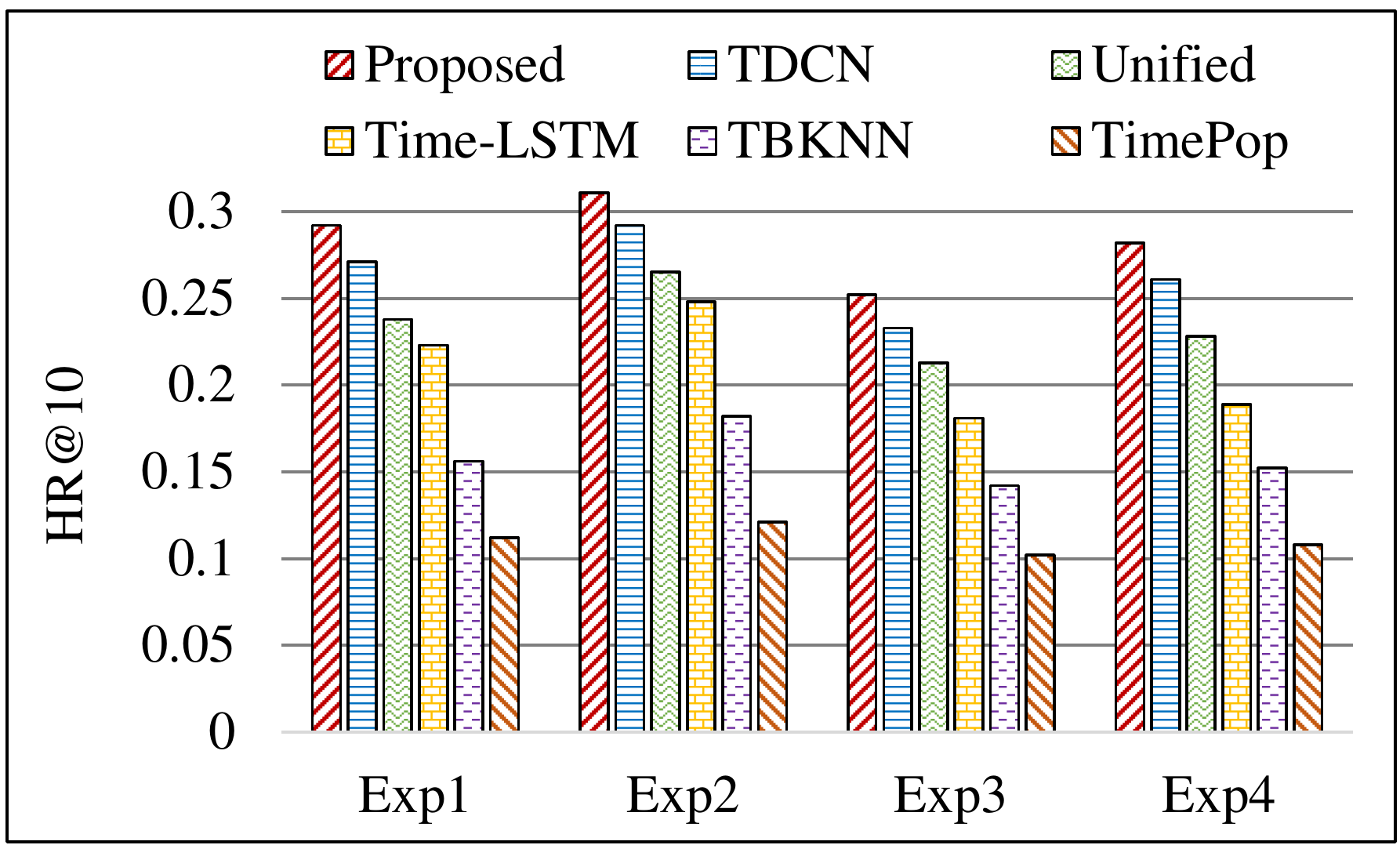}
        \caption{Existing user}
        \label{fig:res_existuser}
    \end{subfigure}\\
    \begin{subfigure}[b]{0.7\linewidth}
        \includegraphics[width=\textwidth]{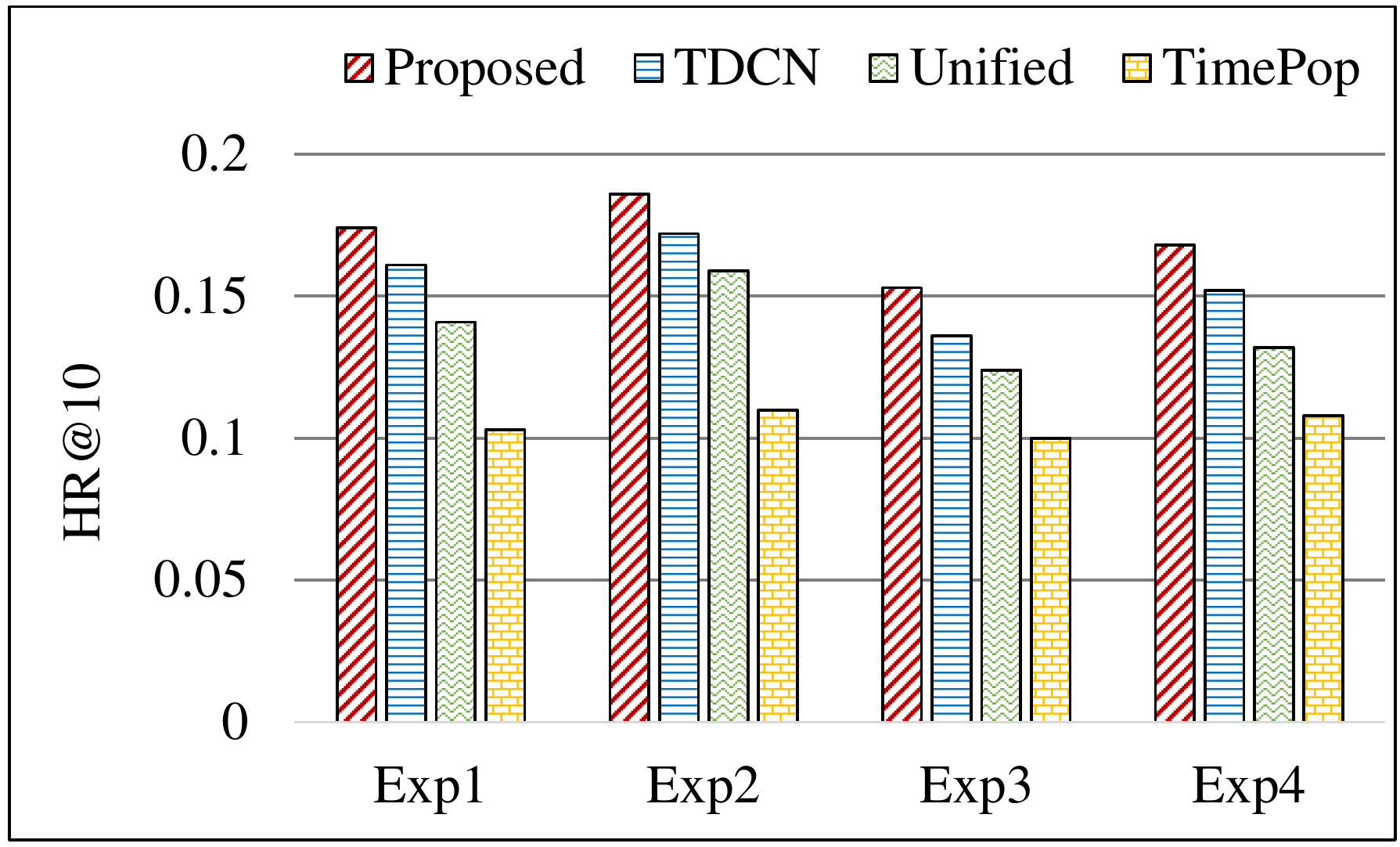}
        \caption{New user}
        \label{fig:res_newuser}
  \end{subfigure}
  \caption{Recommendation accuracy comparisons against baselines.}
\end{figure}

We plot HR@10 under the different experimental settings for existing and new users (see Figures \ref{fig:res_existuser} and \ref{fig:res_newuser}). In general, HR is higher when models are trained in biweekly intervals (Exp. 1 and Exp. 2), compared to monthly intervals (Exp. 3 and Exp. 4). This is intuitive because smaller time intervals capture more recent preferences and finer level preference dynamics over time. Larger training sets (Exp 2. and Exp 4.) also obtained slightly higher accuracy, perhaps due to better trained models. Compared to single-network based models (TimePop, TBKNN and Time-LSTM), cross-network models (Unified, TDCN and Proposed) show higher accuracy, across all experimental settings, since auxiliary information mitigates data sparsity issues. Furthermore, the accuracy improvements of TDCN over the Unified approach shows the benefits of incorporating temporal information. The Proposed model consistently outperforms all baselines in all experimental settings. Note that, only cross-network and TimePop models are able to provide recommendations for new users. The inability to utilize auxiliary information makes single-network baselines nonfunctional for new user recommendations. In general, new user recommender accuracy is lower than existing user accuracy, since new user recommendations only rely on source network information. 
\begin{figure}[h!]
    \centering
    \includegraphics[width=0.8\linewidth]{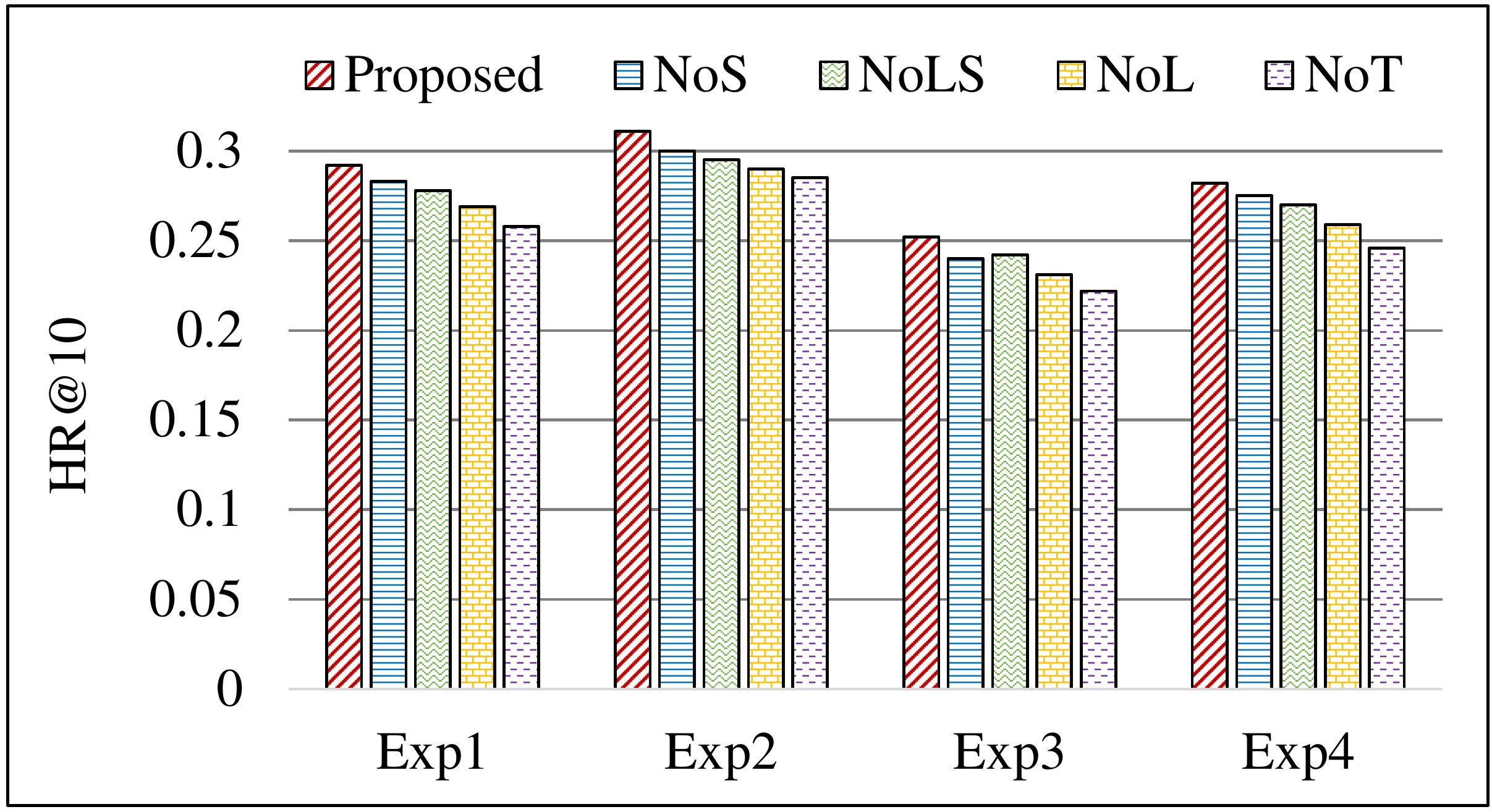}
    \caption{Effects of the proposed short, long and long short term preference components.}
    \label{fig:time_componenets}       
\end{figure}
\begin{figure*}[t!]
    \begin{subfigure}[t]{0.24\linewidth} 
        \centering
        \includegraphics[width=\linewidth]{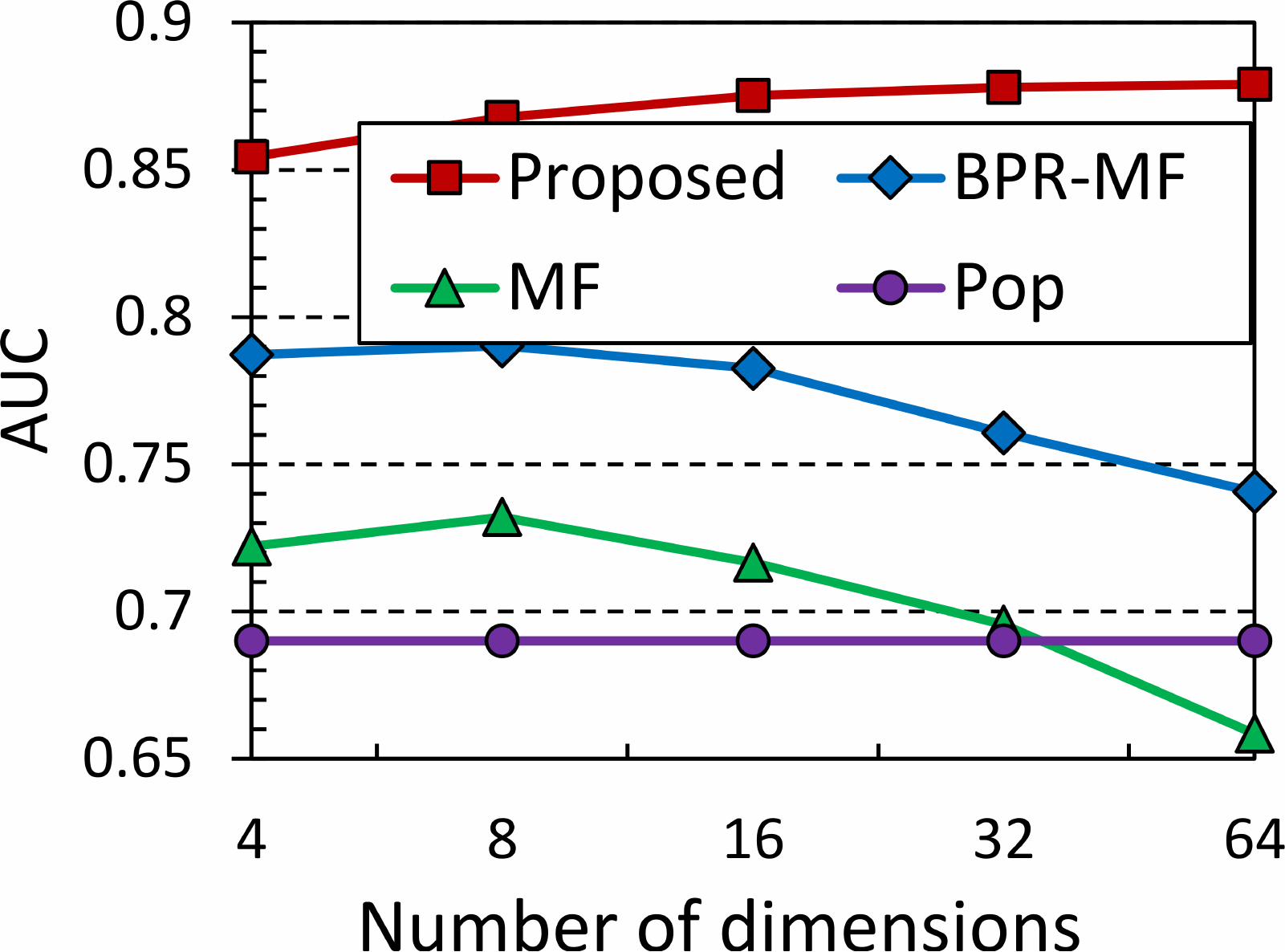}
        \caption{AUC against baseline loss functions.}
        \label{fig:auc_comparisons}
    \end{subfigure}
    \hspace{1pt}
    \begin{subfigure}[t]{0.24\linewidth} 
        \centering
        \includegraphics[width=0.9\textwidth]{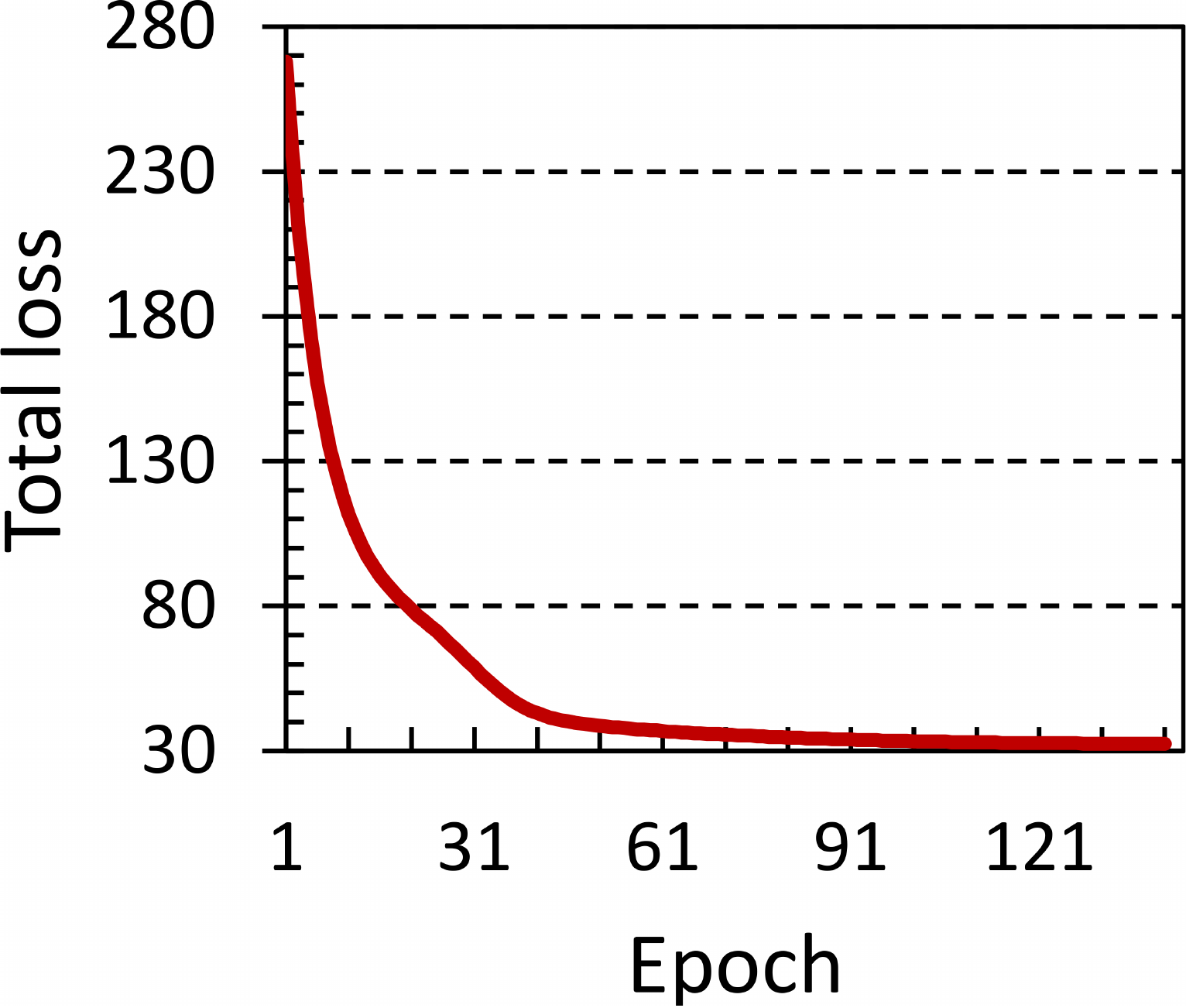}
        \caption{Convergence of listwise loss when optimizing the standard MF model.}
        \label{fig:loss_convergence}
    \end{subfigure}
    \hspace{1pt}
    \begin{subfigure}[t]{0.24\linewidth} 
        \centering
        \includegraphics[width=0.9\textwidth]{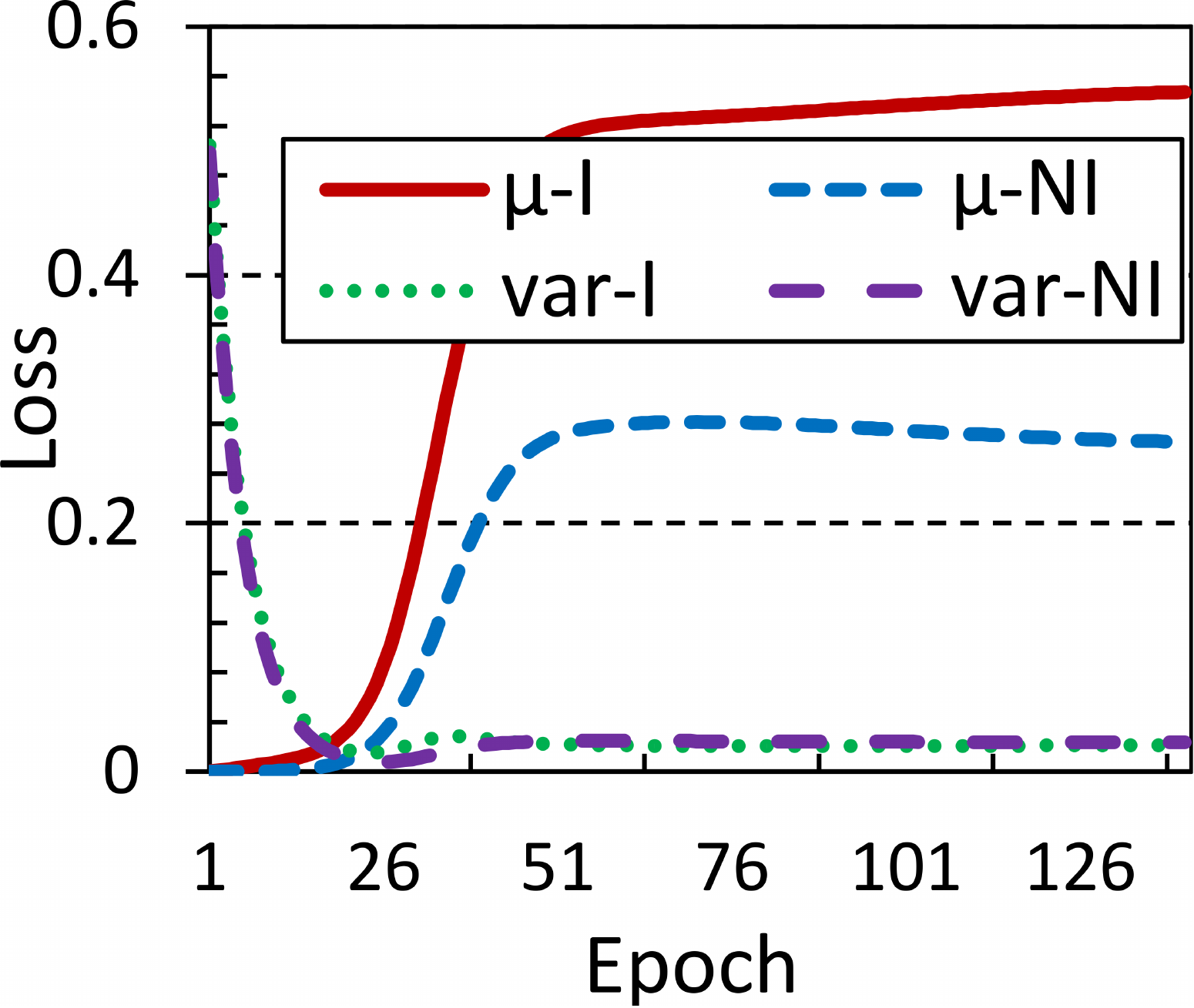}
        \caption{Mean and variance for interacted and non-interacted classes.}
        \label{fig:loss_components}
    \end{subfigure}        
    \hspace{1pt}
    \begin{subfigure}[t]{0.24\linewidth} 
        \centering
        \includegraphics[width=0.9\textwidth]{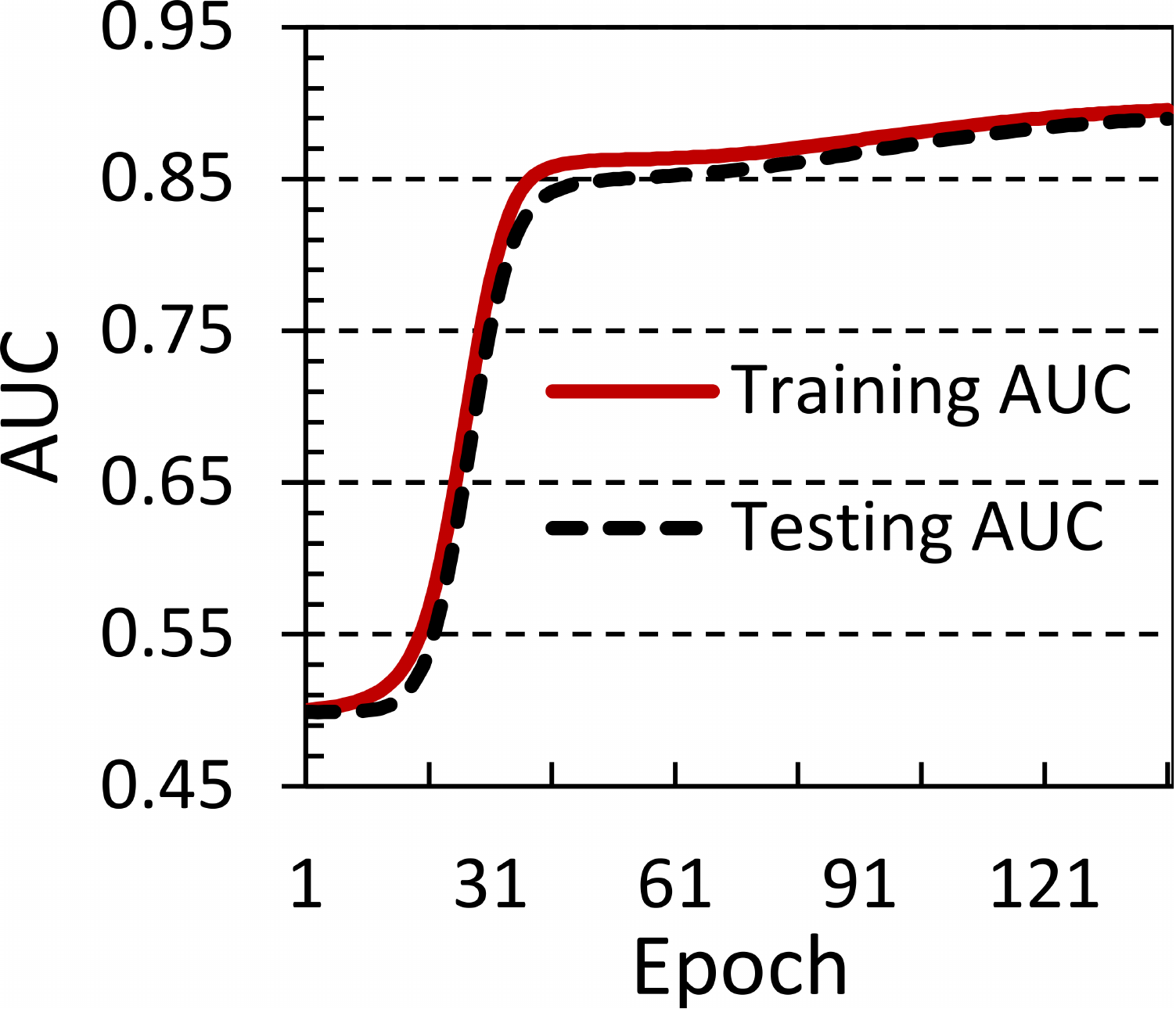}
        \caption{Training and testing AUC during the convergence.}
        \label{fig:auc_train_test}
    \end{subfigure}    
	\caption{Loss function evaluation}\label{fig:loss_function_eval}
\end{figure*}

\subsubsection{Effects of Three Level Temporal Preferences:} To evaluate the effectiveness of short, long and long short term preferences, we removed each component individually and created several variations of the model. The resulting variations NoS, NoLS, NoL and NoT represent baselines \textit{without} short, long short, long and all three components, respectively (see Figure \ref{fig:time_componenets}). Apart from NoT which does not model user preference dynamics, the highest performance drop is attained when the long term preferences component (NoL) is removed. This is intuitive since long term preferences tend to contain the majority of users preferences. Compared to short term preferences, the proposed long short term preferences also have a higher impact on accuracy, which illustrates the effectiveness of the proposed long short term preference component.

\subsubsection{Novelty and Diversity:} Accuracy alone is insufficient to comprehensively evaluate recommender quality. For example, recommending a similar set of interesting items to a user would result in higher accuracy, but he would lose interest in recommendations over time. Therefore, we also measured novelty \cite{zhang2013definition} and diversity \cite{avazpour2014dimensions} of recommended items. We observed that cross-network solutions outperform single network solutions as they exploit diverse user preferences from source networks. The proposed model outperforms the closest TDCN approach by 8.2\% and 9.1\% in novelty and diversity. Further experiments show that compared to long term preferences, the use of short and long short term preferences improved novelty and diversity measures.

\subsection{Loss Function Evaluation}
The proposed listwise loss criterion is generic and can be used with multiple recommendation approaches. However, we optimized MF using the proposed loss function for direct comparisons against the MF based baselines as follows:
\begin{align}
    \hat{r}_{ui} & = \sum_{f=1}^k \boldsymbol{w_{uf}} \cdot \boldsymbol{h_{if}}\in \mathbb{R}\nonumber
\end{align} 
\begin{align}
    L_{lw} & = \sum_{u}\sum_{i\in I^+}\sum_{j \in I\setminus I^+}\big(1-\mu_{ui}\big)^2 + \mu_{uj}^2 + \sigma^2_{ui} + \sigma^2_{uj}\nonumber
\end{align}
where $\hat{r}_{ui}$ is the rating prediction function in MF for a user-item pair, computed by taking the inner product between d-dimensional user and item latent vectors ($\boldsymbol{w_{uf}} \in \mathbb{R}^d$ and $\boldsymbol{h_{if}} \in \mathbb{R}^d$). Note that, the same function is used across all MF based baselines to evaluate the loss functions. The only difference is that we used the proposed $L_{lw}$ loss function to better optimize the $\boldsymbol{w_{uf}}$ and $\boldsymbol{h_{if}}$ factors for the ranking task. As stated above, the baselines use their own popular loss functions (both pointwise and pairwise), instead.

\subsubsection{Performance Comparison:} We compared AUC scores for all models under varying number of dimensions for latent user and item representations ($d$). The non-personalized baseline (Pop) shows the lowest performances. Among the rest of the personalized MF based approaches (Proposed, BPR-MF and MF), pairwise optimizations consistently outperform pointwise optimizations due to their natural support for ranking. However, since none of them are fully optimized for listwise ranking, the proposed listwise optimization criterion consistently outperformed all baselines and is also the least affected by the small variations of $d$ (see Figure \ref{fig:auc_comparisons}).

\subsubsection{Model Convergence:} The total loss converges at around $60$ epochs as the mean values of interacted and non-interacted ratings ($\mu_I$ and $\mu_{NI}$) approach their maximum and minimum values (see Figures \ref{fig:loss_convergence} and \ref{fig:loss_components}). Simultaneously, the variance values for the corresponding classes ($\sigma_I$ and $\sigma_{NI}$) lead to 0 at early epochs resulting in very low inter-class distances. After the overall loss reaches its minimum, the mean value of the interacted class slowly increases, while the mean value of the non-interacted class slowly decreases. At this stage, the model is still being marginally optimized, and it is visible from the continuous increase in AUC (see Figure \ref{fig:auc_train_test}).

\section{Conclusion and Further Work}
Typical recommender systems fail to conduct recommendations for both new and existing users by considering the dynamic nature of user preferences. In addition, existing implicit feedback based systems are not fully optimized to rank a list of items. Thus to the best of our knowledge, we proposed the first time aware unified cross-network recommender solution with a generic listwise loss function. The proposed model first transfers and integrates user interactions from multiple networks in a cross-network topical layer. Second, captures dynamic user preferences in three temporal levels, namely short, long and long short term preferences. The proposed deep learning model directly utilized the computed source network user preferences for new user recommendations and for existing users, they are integrated with target network preferences. We used two datasets, CrossNet and MovieLens to evaluate the proposed model and the listwise optimization criterion and they consistently outperformed multiple state-of-the-art baselines. As future work, the proposed model can be extended with social (e.g.,  follower and followee) and item information. The intuition for the proposed listwise optimization criterion can also be extended for ranking under explicit feedback. Overall, we believe the paper makes a significant contribution to the cross-network recommender field and the use of implicit feedback.
\section{ Acknowledgments}
This research has been supported by the Singapore Ministry of Education Academic Research Fund Tier 2 under MOE’s official grant number MOE2018-T2-1-103. We gratefully acknowledge the support of NVIDIA Corporation for the donation of the Titan Xp GPU used in this research.

\bibliographystyle{aaai}
\small
\bibliography{aaai20ref}
\end{document}